\documentclass[a4paper]{jpconf}
\usepackage{graphicx}

\usepackage{amsmath}
\usepackage{amssymb}
\usepackage{float}
\usepackage[dvips]{color}
\usepackage{colordvi}
\usepackage{url}

\begin{document}
\begin{flushright}
TIFR/TH/11-49\\
CCTP-2011-39
\end{flushright}

\title{What is the gravity dual of \\
the confinement/deconfinement transition \\
in holographic QCD?}

\author{Gautam Mandal}

\address{Department of Theoretical Physics, Tata Institute of Fundamental Research, Mumbai 400 005, India}

\ead{mandal@theory.tifr.res.in}

\author{Takeshi Morita}

\address{Crete Center for Theoretical Physics, Department of Physics, University of Crete,
71003 Heraklion, Greece}

\ead{takeshi@physics.uoc.gr}

\begin{abstract}

We study the gravity dual of four dimensional pure Yang-Mills theory through D4 branes, as proposed by Witten (holographic QCD).
 In this holographic QCD, it has been widely believed that the confinement phase in the pure Yang-Mills theory corresponds to the  solitonic D4 brane in gravity and the deconfinement phase corresponds to the black D4 brane.  
We inspect this conjecture carefully and show that the correspondence between the black D4 brane and the deconfinement phase is not correct. 
Instead, by using a slightly different set up, we find an alternative gravity solution called ``localized soliton'', which would be properly related to the deconfinement phase.
In this case, the confinement/deconfinement transition is realized as a Gregory-Laflamme
 type transition.  
We find that our proposal naturally explains several known properties of QCD.

\end{abstract}

\def\eq#1{(\ref{#1})}

\section{Introduction}

In this letter\footnote{This letter is based on the proceedings of the Seventh International Conference `Quantum Theory and Symmetries' held in Prague 2011 and XVII European Workshop
on String Theory 2011 held in Padua. 
The talk is based on our recent work \cite{Mandal:2011ws}.}, we will focus on holographic QCD from D4 branes \cite{Witten:1998zw, Aharony:1999ti, Gross:1998gk, 
  Kruczenski:2003uq, Sakai:2004cn}. 
  We will discuss some
problems  with
the usual correspondence between the
confinement/deconfinement transition in QCD and the Scherk-Schwarz
transition--- between a solitonic D4 brane and a black D4 brane--- in the gravity dual.
Some of these problems were first discussed in \cite{Aharony:2006da}.
We will specifically show that the black D4 brane cannot be identified
with the (strong coupling continuation of the) deconfinement phase in QCD
in four dimensions. 
As a resolution of these problems, we will
propose an alternative scenario in which the confinement/deconfinement
transition corresponds to a Gregory-Laflamme transition
\cite{Gregory:1994bj,hep-th/0204047, Aharony:2004ig, Harmark:2004ws, hep-th/0409111, hep-th/0411240, arXiv:0706.0188} between a uniformly distributed soliton and
a localized soliton in the IIB frame.
The scenario we propose suggests that we need to reconsider several
previous results in holographic QCD including phenomena
related to the Sakai-Sugimoto model
\cite{Sakai:2004cn}. 

\section{Holographic construction of QCD}

In this section, we will review the construction of four dimensional
$SU(N)$ pure Yang-Mills theory (YM4) from $N$ D4 branes
\cite{Witten:1998zw}.  Let us first consider a 10 dimensional
Euclidean spacetime with an $S^1$ and consider D4 branes wrapping on
the $S^1$. 
We define the coordinate along this $S^1$ as $x_4$ and its periodicity as $L_4$.
The effective theory on this brane is a 5 dimensional supersymmetric Yang-Mills theory (SYM5) on the $S^1_{L_4}$.
For the fermions on the brane, the boundary condition along the circle can
be AP (antiperiodic) or P (periodic); to specify the theory, we must
pick one of these two boundary conditions. Let
us take the AP boundary condition.  This gives rise to fermion masses
proportional to the Kaluza-Klein scale $1/L_4$, leading to
supersymmetry breaking (this is called the SS --- Scherk-Schwarz---
mechanism).  This, in turn, induces masses for the adjoint scalars and for
$A_4$, which are proportional to $\lambda_4/L_4$ at
one-loop. 
Therefore, if $\lambda_4$ is sufficiently small and the
dynamical scale $\Lambda_{YM}$ and temperature are much less than both the above mass
scales, then the fermions, adjoint scalars and KK modes are decoupled and the 5 dimensional supersymmetric Yang-Mills theory is reduced to a four dimensional pure Yang-Mills theory.
The precise conditions to obtain YM4 from  SYM5 are
\begin{align} 
\lambda_4 \ll 1 , \quad \beta \gg \frac{L_4}{\lambda_4}.
\label{gauge-cond}
\end{align}

By taking the large $N$ (Maldacena) limit of this system \cite{Itzhaki:1998dd} at low temperatures, we obtain the dual gravity description of the compactified 5 dimensional SYM theory \cite{Witten:1998zw}, which consists, at low temperatures, of  a solitonic D$4$ brane solution wrapping the
$S^1_{L_4}$.
  The explicit metric is given by 
\begin{align}
ds^2 =& \alpha' \left[\frac{u^{3/2}}{\sqrt{d_4 \lambda_5 }}
\left( dt^2  + \sum_{i=1}^{3} 
dx_i^2+f_4(u) dx_4^2 \right)+ \frac{\sqrt{d_4 \lambda_5 }}{u^{3/2}}\left(  
\frac{du^2}{f_4(u)} 
+ u^2 d\Omega_{4}^2 \right)  \right], \nonumber \\
& f_4(u)=1-\left( \frac{u_0}{u}\right)^3 , \quad e^{\phi}=\frac{\lambda_5}{(2\pi)^2N}
\left(\frac{u^{3/2}}{\sqrt{d_5 \lambda_5}}  \right)^{1/2}, \quad u_0=\frac{\pi \lambda_4}{9 L_4} ,  
 \label{metric-SD4}
\end{align}
where $\lambda_5= \lambda_4 L_4 $ and $d_4=1/4\pi$.
 This gravity solution is not
always valid. E.g. in order that the stringy modes can be ignored, we
should ensure that the curvature in string units must be small \cite{Itzhaki:1998dd}.
 This condition turns out to be equivalent to
\begin{align} 
\lambda_4 \gg 1 .
\label{gravity-cond-SD4}
\end{align} 
This is opposite to the condition (\ref{gauge-cond}).
Thus, the gravity solution can describe SYM5 but cannot directly describe YM4.
This is a common problem in the construction of holographic duals of non-supersymmetric gauge theories.
However, in principle, leading order effects of stringy modes can be 
computed by perturbatively evaluating $1/\lambda_4$ corrections in 
the gravity theory and, as in 
case of the standard strong coupling expansion
in gauge theory (see, e.g. \cite{Drouffe:1978dn}), we may obtain appropriate weak coupling results,
{\em provided no phase transition occurs between the weak and strong coupling regimes}.
Although the gravity solution (\ref{metric-SD4}) is just the leading term of such an expansion, we should be able to infer
qualitative properties of the gauge theory by extrapolating.
Many interesting results, including the qualitative predictions in \cite{Witten:1998zw, Aharony:1999ti, Gross:1998gk, Sakai:2004cn}, have been obtained using this prescription. We will proceed to describe the thermodynamics of Yang-Mills theory 
from gravity in this spirit.

\section{Phase structure of the dual gravity}

In this section we investigate the thermodynamic phase structure of the gravity
solutions and compare the result with the phase structure of pure Yang-Mills theory.
To discuss holographic QCD at finite temperatures, we begin by compactifying Euclidean time in the boundary theory on a circle with periodicity $\beta=1/T$. 
In order to determine the gravitational theory, we need to fix the
periodicity of fermions in the gauge theory along the time cycle. 
Let us recall that the gauge theory
of interest here is pure Yang Mills theory in four dimensions,
which does not have fermions.  Fermions reappear when the validity
condition of the gravity description (\ref{gravity-cond-SD4}) is enforced and $\Lambda_{YM}$ goes above
the KK scale $1/L_4$. 
In this sense, fermions are an artifact of the
holographic method and in the region of validity of the pure YM
theory, the periodicity of the fermion should not affect the gauge
theory results.

Indeed we can obtain the thermal partition function of  YM4 from  SYM5 on $S^1_\beta \times S^1_{L_4} $ with either the (AP,AP) or (P,AP) boundary condition\footnote{\label{ftnt-top} Here and elsewhere the boundary
  conditions will always refer to those of the boundary theory along
  $S^1_\beta \times S^1_{L_4}$, respectively.}:
\begin{align} 
Z_{\rm (AP,AP)}^{{\rm SYM}5}&=\Tr  e^{-\beta H_{ {\rm SYM}5}} 
\to
\Tr e^{-\beta H_{ {\rm YM}4}}, \quad &(\lambda_4 \to 0, \frac{L_4}{\lambda_4 \beta} \to 0), \nonumber \\
Z_{\rm (P,AP)}^{{\rm SYM}5}&=\Tr (-1)^F e^{-\beta H_{ {\rm SYM}5}} 
\to
\Tr e^{-\beta H_{ {\rm YM}4}}, \quad &(\lambda_4 \to 0, \frac{L_4}{\lambda_4 \beta} \to 0).
\label{4d-limit}
\end{align} 
where the limit ensures that the fermions and adjoint scalars are
decoupled because of large mass.  Since we
recover pure YM4 in both cases, it is pertinent to
study gravity solutions corresponding to both these boundary conditions.

The gravity solutions appearing in the (AP,AP) case  \cite{Horowitz:1998ha} are summarized as
\begin{align} 
\begin{tabular}{l|c|c}
Low temperature & Solitonic D4 solution &  $W_0=0$, $W_4 \neq 0$   \\
\hline 
High temperature & Black D4 solution &  $W_0 \neq 0$, $W_4 = 0$  \\
\end{tabular}
\label{table-APAP}
\end{align} 
Here the metric of the black D4 solution is given by
\begin{align}
ds^2 =& \alpha' \left[\frac{u^{3/2}}{\sqrt{d_4 \lambda_5 }}
\left( f_4(u)dt^2  + \sum_{i=1}^{3} 
dx_i^2+ dx_4^2 \right)+ \frac{\sqrt{d_4 \lambda_5 }}{u^{3/2}}\left(\frac{du^2}{f_4(u)} 
+ u^2 d\Omega_{4}^2 \right)  \right], ~~ u_0=\frac{\pi \lambda_4 L_4}{9\beta^2}  
\label{metric-BD4}
\end{align}
Note that this metric is related to the solitonic D4 brane (\ref{metric-SD4}) through the $Z_2$ symmetry: $\beta \leftrightarrow L_4$.
Thus, the free energies of these solutions are coincident at the self-dual point $\beta=L_4$ and a phase transition called  Scherk-Schwarz transition happens there.
In (\ref{table-APAP}), $W_0$ and $W_4$ are the Polyakov loop operators:
\begin{align} 
 W_0= \frac1N {\rm Tr } P e^{i \int_0^{\beta} A_0 dx^0 },
 \quad
 W_4= \frac1N {\rm Tr } P e^{i \int_0^{L_4} A_4 dx^0 }.
\label{Polyakov}
\end{align} 
Vacuum expectation values of these operators characterize the phases of the gauge theory. These also act as order parameters for the $Z_N \times Z_N$ `centre symmetry'
(see \cite{Mandal:2011ws} for details) and their vanishing is related to 
 large-$N$ volume independence \cite{Mandal:2011ws, Eguchi:1982nm, Gocksch:1982en}.

The gravity solutions appearing in the (P,AP) case are
\begin{align} 
\begin{tabular}{l|c|c}
Low temperature & Solitonic D4/uniformly smeared solitonic D3 &  $W_0=0$, $W_4 \neq 0$   \\ 
\hline
High temperature & Localized solitonic D3 solution &  $W_0 \neq 0$, $W_4 \neq 0$  \\
\end{tabular}
\end{align} 
Note that we need to take a T-dual along the temporal $S^1_\beta$ circle and go to the IIB frame to see the high temperature region, since the masses of the winding modes of the IIA string wrapping along the temporal circle become light when $T > \sqrt{\lambda_4}/L_4$.
This T-dual maps the solitonic D4 solution in the IIA to solitonic D3 branes uniformly smeared along the dual temporal circle in the IIB, with
metric given by
\begin{align}
ds^2 =& \alpha' \left[\frac{u^{3/2}}{\sqrt{d_4 \lambda_5 }}
\left( \sum_{i=1}^{3} 
dx_i^2+f_4(u) dx_4^2 \right)+ \frac{\sqrt{d_4 \lambda_5 }}{u^{3/2}}\left(  \frac{du^2}{f_4(u)} + d{t'}^2
+ u^2 d\Omega_{4}^2 \right)  \right].
\label{metric-SSD3}  
\end{align}
Here $ \alpha' t'$ is the dual of $t$, hence $t'$ has a periodicity $\beta'=(2\pi)^2/\beta=(2\pi)^2T$.  
This uniform  configuration becomes meta-stable above a critical temperature $\sim L_4/\lambda_4$, and the solitonic D3 branes get localized on the dual temporal circle.
This phase transition is a  Gregory-Laflamme (GL) type transition \cite{Gregory:1994bj,hep-th/0204047, Aharony:2004ig, Harmark:2004ws, hep-th/0409111, hep-th/0411240, arXiv:0706.0188}.
The metric of the localized solitonic D3 brane is approximately described in \cite{Mandal:2011ws}.

The phases in the 4 dimensional pure Yang-Mills theory are
\begin{align} 
\begin{tabular}{l|c|c}
Low temperature & Confinement phase &  $W_0=0$, $W_4 \neq 0$   \\ 
\hline
High temperature & Deconfinement phase &  $W_0 \neq 0$, $W_4 \neq 0$  \\
\end{tabular}
\end{align} 
Here $W_4$ makes an appearance because YM4 is viewed as a limit
\eq{gauge-cond} of SYM5. That $W_4 \neq 0$ follows from the fact that
KK reduction from SYM5 to YM4 only works in this phase; in the phase
with vanishing $W_4$ the thermodynamics is independent of $L_4$
(through the large $N$ volume independence mentioned above) and cannot
describe the KK reduction. Note, further, that the above table works
for either P or AP boundary condition for the SYM5 fermions along the
temporal circle, as we argued in (\ref{4d-limit}).

Let us now compare this phase structure in Yang-Mills theory with
the one found above for gravity.
In the low temperature regime, the properties of the Polyakov loop operators in the solitonic D4 brane agree with those of the confinement phase.
Thus these two phases may be identified.
In the high temperature regime, it is the localized D3 soliton 
which matches the deconfinement phase in terms of properties of $W_0$ and $W_4$ whereas the black D4 brane solution does not. This, in particular, means that
the $Z_N \times Z_N$ symmetry is realized in the same was in the localized 
D3 soliton phase as in the deconfinement phase, whereas it is realized differently
in the black D4 solution.
Thus the localized D3 solution may correspond to the deconfinement phase but the black D4 brane does not.

\begin{figure}
\begin{center}
\includegraphics[scale=.6]{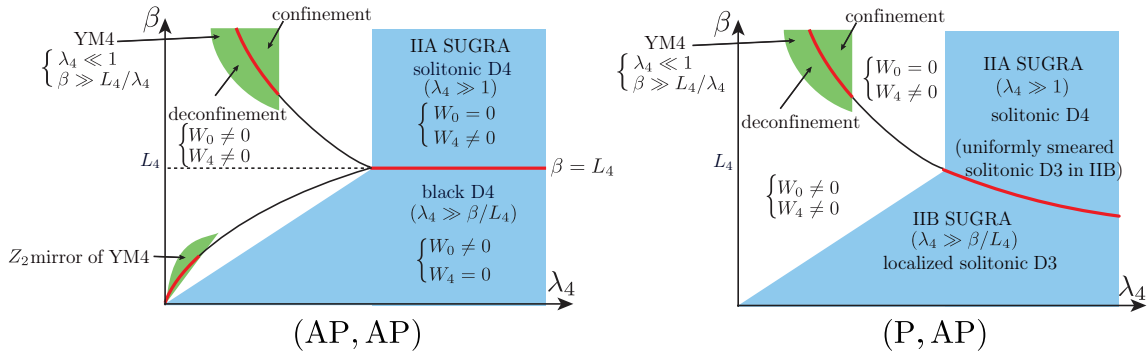}
\caption{The phase structures of the five dimensional SYM on $S^1_\beta \times S^1_{L_4} $
  with the (AP,AP) and (P,AP) boundary condition.
  The gravity analysis is valid in the strong coupling region (the blue region).
The 4 dimensional YM description is valid in the upper green region, see (\ref{gauge-cond}).
The lower green region in the (AP,AP) case is the mirror of the upper one via the $Z_2$ symmetry $\beta \leftrightarrow L_4$. 
The solid black lines correspond to a minimal extrapolation of the phase boundaries through the intermediate region. The dotted line denotes another possible phase transition which is allowed by the $Z_2$ symmetry.
}
\label{fig-D4-phase}
\end{center}
\end{figure}

Indeed the expected phase structures of  SYM5 with the (AP,AP) and (P,AP) boundary condition are shown in Figure \ref{fig-D4-phase}.
In the (AP,AP) case, at least one phase transition has to occur between the black D4 brane solution and the deconfinement phase.
(This is also expected through the $Z_2$ symmetry: $S^1_\beta \leftrightarrow S^1_{L_4} $ \cite{Aharony:2006da}. See also \cite{Aharony:2005ew}.)
Therefore the previous conjecture that the black D4 brane corresponds to the deconfinement phase of the Yang-Mills theory is not correct, since these two
phases are not smoothly connected (see the remarks below \eq{gravity-cond-SD4}
in {\em italics}).

By contrast, in the (P,AP) case, the gravity solutions in the strong coupling regime may smoothly continue to the phases of the weakly coupled 4 dimensional Yang Mills theory.
Thus we propose that we should look at these gravity solutions in the (P,AP) case to investigate the dynamics of the Yang-Mills theory.
Especially the GL transition between the smeared  D3 soliton and the localized D3 soliton would correspond to the confinement/deconfinement transition in the Yang-Mills theory\footnote{Ref. \cite{Catterall:2010fx} considers lattice SYM2 arising from D1 branes compactified on (P,AP) circles, and obtains a phase diagram  similar to the one in Figure \ref{fig-D4-phase} where D4 branes are replaced by D1 branes.
This provides support for the proposal presented in this paper.}.

\section{Gregory-Laflamme transition as a Hagedorn transition}

Our proposal opens up the interesting possibility of a relation
between the GL transition and the Hagedorn transition.
It is known that the GL instability is an instability of the 
KK modes of the graviton along the compact circle \cite{Gregory:1994bj}
(such modes develop an imaginary frequency and grow exponentially).
In our case, the KK modes along the dual
temporal circle, which cause the GL instability in the IIB
description, would be mapped to winding modes around the temporal circle
through the T-duality \cite{Harmark:2007md, Ross:2005vh}.  
This indicates that the GL transition is associated with the
tachyonic instability of the winding modes of the IIA string.  
This is similar to the Hagedorn transition in string theory
\cite{Atick:1988si}, where the temporal winding modes cause the
instability.  
Thus the GL transition in the IIB description might
correspond to the Hagedorn transition in the IIA description.

Note that on the gauge theory side also, the
confinement/deconfinement transition has been shown to be related to
the Hagedorn transition \cite{Sundborg:1999ue, Aharony:2003sx}.  This
makes it plausible that the Hagedorn transition in the Yang-Mills
theory continues to the Hagedorn transition in the IIA string, which,
as we argued above, is possibly the dual of the GL transition in the
IIB supergravity.

\section{Conclusion}

In this study, we showed that the identification between the deconfinement phase and the black D4 brane solution in the (AP,AP) boundary condition is not correct and proposed a resolution by using the (P,AP) boundary condition.
This result suggests that we need to reconsider several
previous results, in which the black D4 brane was employed in high
temperature holographic QCD including the Sakai-Sugimoto model \cite{Sakai:2004cn}.
  One important ingredient in the Sakai-Sugimoto
model is the mechanism of chiral symmetry restoration at high
temperatures \cite{Aharony:2006da}.  
In \cite{Mandal:2011ws}, we proposed a new mechanism
for chiral symmetry restoration in our framework.
Another important issue is to explore what geometry corresponds to the deconfinement phase in the real time formalism; this is important, e.g., for the investigation of the viscosity ratio.

Problems similar to the above had also been encountered in the study of two dimensional bosonic gauge theory in \cite{Mandal:2011hb}. 
This indicates that the issues addressed in this paper are rather general in the discussion of holography for non-supersymmetric gauge theories at finite temperatures.

\section*{Acknowledgments}
We would like to thank Ofer Aharony, Avinash Dhar, Saumen Datta, Koji Hashimoto, Yoshimasa Hidaka,
Elias Kiritsis, Matthew Lippert, Shiraz Minwalla, Rene Meyer, Rob Myers, Vasilis Niarchos,
Tadakatsu Sakai, Shigeki Sugimoto and Tadashi Takayanagi for valuable discussions and comments. 
T.M. is partially supported by Regional Potential program of the
E.U. FP7-REGPOT-2008-1: CreteHEPCosmo-228644 and by Marie Curie
contract PIRG06-GA-2009-256487.

\section*{References}


\end{document}